\pdfoutput=1
\documentclass[
preprint,
 amsmath,amssymb,
aps,
prl,
endfloats*,
]{revtex4-1}

\usepackage{color}
\usepackage{rotating}
\usepackage{lipsum}
\usepackage{graphicx}
\usepackage{dcolumn}
\usepackage{bm}
\usepackage{textcomp}
\usepackage{changes}
\setremarkmarkup{(#2)}
\def\andname{\hspace*{-0.5em}}

\begin{document}

\title{Band engineering of a magnetic thin film rare earth monopnictide}

\author{Hisashi Inoue$^{1}$\footnote{These authors contributed equally to the work.}, Minyong Han$^{1{\rm a}}$, Mengli Hu$^{2{\rm a}}$, Takehito Suzuki$^{1}$, Junwei Liu$^{2}$\footnote{liuj@ust.hk}, Joseph G. Checkelsky$^{1}$\footnote{checkelsky@mit.edu}}
\affiliation{$^{1}$Department of Physics, Massachusetts Institute of Technology, Cambridge, Massachusetts 02139, USA}
\affiliation{$^{2}$Department of Physics, Hong Kong University of Science and Technology, Clear Water Bay, Hong Kong, China}

\date{\today}

\begin{abstract}
\textbf{
Realizing quantum materials in few atomic layer morphologies is a key to both observing and controlling a wide variety of exotic quantum phenomena.  This includes topological electronic materials, where the tunability and dimensionality of few layer materials have enabled the detection of $Z_{2}$, Chern, and Majorana phases.  Here, we report the development of a platform for thin film correlated, topological states in the magnetic rare-earth monopnictide ($RX$) system GdBi synthesized by molecular beam epitaxy.  This material is known from bulk single crystal studies to be semimetallic antiferromagnets with Neel temperature $T_{N} = 28$ K and is the magnetic analog of the non-$f$-electron containing system LaBi proposed to have topological surface states.  Our transport and magnetization studies of thin films grown epitaxially on BaF$_{2}$ reveal that semimetallicity is lifted below approximately 8 crystallographic unit cells while magnetic order is maintained down to our minimum thickness of 5 crystallographic unit cells. First-principles calculations show that the non-trivial topology is preserved down to the monolayer limit, where quantum confinement and the lattice symmetry give rise to a $C=2$ Chern insulator phase.  We further demonstrate the stabilization of these films against atmospheric degradation using a combination of air-free buffer and capping procedures. These results together identify thin film $RX$ materials as potential platforms for engineering topological electronic bands in correlated magnetic materials.
}
\end{abstract}

\maketitle

\textbf{Introduction}

The rare-earth monopnictides ($RX$, where $R$ and $X$ denote the rare-earth and pnictogen elements) are a class of materials which host a rich variety of magnetic and electronic phases ranging from ferromagnetic semiconductors to antiferromagnetic semimetals \cite{Duan2007,Wu2017}. Recent studies have shown that for $X = $ Bi the large spin-orbit coupling results in band inversion and topologically protected surface states \cite{Zeng2015,Nayak2017,Lou2017,Feng2018,Duan2018,Kuroda2018,Alidoust2016,Kane2005,Qi2011,Bernevig2006}.  Given the strong coupling between the conduction electrons and localized magnetic moments of $RX$ systems, this offers the opportunity to study the non-trivial electronic topology of correlated electrons \cite{Mong2010,Liu2016,Chang2013,Winnerlein2017,Liu2008}.  A challenge for these systems is the significant semimetallic band overlap of the conduction and valence bands (described below), which for transport studies precludes the isolation of the surface response.  However, while realization of few monolayer morphologies has proven to be a powerful method to remove parasitic bulk conductance from unintentional doping in conventional three-dimensional topological insulators (TIs) \cite{Zhang2010,Zhang2011,Checkelsky2011,Li2010}, theoretical calculations of $RX$ systems in this limit suggest further that quantum confinement has the potential to energetically isolate the topologically non-trivial bands \cite{Li2015}.

\begin{figure}
\includegraphics[width=12cm]{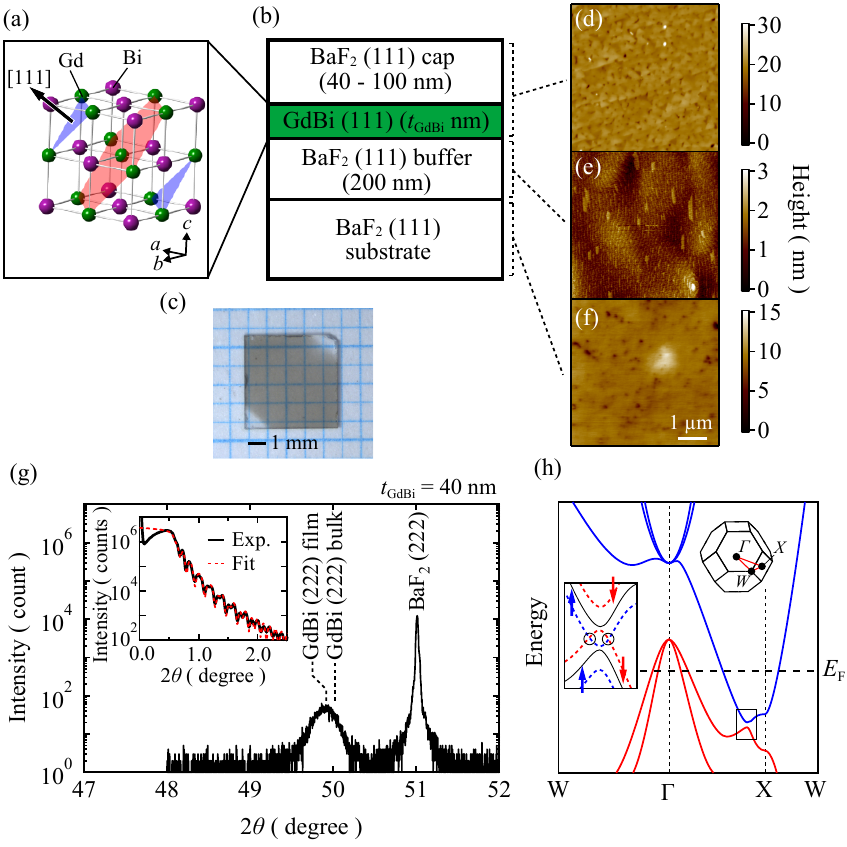}
\caption{\label{fig1}
Basic characterization of GdBi thin films. (a) Crystal structure of antiferromagnetic GdBi. The shaded areas denote the (111) planes, where spins are ferromagnetically aligned. The spin orientations are opposite between the red and blue shaded areas. (b) Schematic and (c) optical microscope image of GdBi thin film structure. (d)(e)(f) Atomic force microscope image of (d) the BaF$_2$ and AlN cap layers, (e) the BaF$_2$ buffer layer, and (f) the surface of an annealed BaF$_2$ (111) substrate. (g) X-ray diffraction data measured on film with $t_{\rm GdBi} = 40$ nm. Inset: X-ray reflectometry data measured on the same sample. The red curve is a fit result to a model structure (see text). (h) Schematic band structure of bulk GdBi. Red and blue curves denote the valence and conduction bands, respectively. The upper inset shows the Brillouin zone and high symmetry lines corresponding to the band dispersion shown in the main panel.  The lower insets shows an expanded view of the band structure around conduction and valence band anticrossing. Exchange splitting (shown in the dashed curve) may push the conduction and valence bands with opposite spins toward each other to generate Weyl nodes (marked with the circles).
}
\end{figure}

Here we report the synthesis and study of epitaxial thin films of the $R = $ Gd compound GdBi and show that it can be engineered toward an insulator by two-dimensional (2D) quantum confinement while retaining its magnetic and topological properties.  As shown in Fig$.$ \ref{fig1} (a), GdBi has a rock-salt structure (space group Fm$\overline{3}$m) and hosts type-II antiferromagnetic (AFM) order below $T_{\rm N} = 28$ K \cite{Duan2007,Ye2018} with Gd moments ferromagnetically aligned within the $\left\{111\right\}$ planes and antiferrromagnetically stacked along the $\left\langle 111 \right\rangle$ directions \cite{Duan2006}.  The electronic band structure in the vicinity of the Fermi level $E_{\rm F}$ consists of a Gd $5d$-derived conduction band around the $X$ points and two Bi $6p$-derived valence bands around the $\mathit{\Gamma}$ point (see Fig$.$ \ref{fig1} (h)). These bands overlap by approximately 1 eV, thus constituting a semimetal. Midway in momentum between $\mathit{\Gamma}$ and $X$ a band crossing occurs, which when hybridized by spin-orbit coupling is proposed to give rise to a topologically non-trivial gap \cite{Zeng2015}. This spin-orbit induced gap also offers a possibility for realizing Weyl nodes in the magnetic phase of GdBi in an applied field, taking advantage of strong exchange splitting in a canted configuration (Fig$.$ \ref{fig1} (h) inset) \cite{Hirschberger2016,Suzuki2016,Cano2017,Armitage2018}.

While photoemission studies of bulk single crystals have proven instrumental for determining the bulk and surface electronic structure of the $RX$ systems \cite{Zeng2016a,Nayak2017,Kuroda2018,Lou2017,Feng2018,Wu2017,Alidoust2016}, as noted above it is of significant interest to realize few monolayer morphologies of these materials to study their transport properties as well as enable \textit{in-situ} control via electrostatic gates.  
Given their cubic rock-salt structure, this is most appropriately achieved by thin film growth, which we employ here via molecular beam epitaxy (MBE).  Subsequent structural, electrical transport, and magnetic characterization demonstrate the high quality of these materials.  First principle band structure calculations confirm that nontrivial topology remains upon confinement and leads to a $C=2$ Chern insulator phase in the monolayer limit.  These suggest that thin film GdBi may be an ideal platform to investigate correlation between magnetism and topologically non-trivial surface states \cite{Mong2010}.\\

\textbf{Fabrication of GdBi epitaxial thin films by MBE}

High quality epitaxial thin films of GdBi $(111)$ were synthesized by MBE.  Single crystalline BaF$_2$ $(111)$ is used for the substrate, which has a cubic lattice constant $a_{\rm BaF2} = 0.620$ nm well-matched with that of GdBi $a_{\rm GdBi} = 0.632$ nm. Prior to GdBi deposition, 200 nm of an epitaxial BaF$_2$ $(111)$ buffer layer was deposited to improve surface flatness.  As shown in Fig$.$ \ref{fig1} (e) and (f), atomic force microscope (AFM) images of the BaF$_2$ buffer layer show a smooth surface with atomically flat terraces and steps with height corresponding to the spacing between successive (111) planes of BaF$_2$. The GdBi $(111)$ layer was grown at temperature $T = 400$ $^\circ$C with the thickness varied from $t_{\rm GdBi} = 5$ nm to 40 nm. Finally, the structure was capped with 40 to 100 nm of epitaxial BaF$_2$ (see Methods for details). The overall structure is summarized in Fig$.$ \ref{fig1} (b) with an optical photograph of a typical film shown in Fig$.$ \ref{fig1} (c). For a magnetic characterization by SQUID magnetometer and for structural characterizations, samples were capped with an additional AlN layer using atomic layer deposition.\\

\textbf{Structural characterization}

Growth of single crystalline GdBi was confirmed by X-ray diffraction as shown in Fig$.$ \ref{fig1} (g), where the peak at $2\theta = 49.90^\circ$ is identified as GdBi $(222)$. A slight shift of this peak from the bulk reference position $2\theta_{\rm ref} = 50.02^\circ$ implies an epitaxial lattice strain of approximately 0.2\%. The AFM image (Fig$.$ \ref{fig1} (d)) and the X-ray reflectivity (XRR) oscillations (Fig$.$ \ref{fig1} (g) inset) confirm a smooth surface on the top BaF$_2$ and AlN cap layers. The thickness of the GdBi layer was calibrated by fitting the XRR oscillations to a model structure simulation (see supplementary materials). We found that the GdBi thin films are extremely sensitive to oxygen and moisture; air exposure of the samples degrades their magnetic and electrical properties within a time scale of seconds even with a BaF$_2$ capping layer (see supplementary materials). Therefore, we added an additional coating of non-aqueous liquid after direct transfer of the film from ultrahigh vacuum to an inert Ar atmosphere for magnetic torque and electrical transport measurements. This allows for the films to preserve their intrinsic physical properties for an adequate amount of time for transfer to characterization apparatus ( $>$ 10 minutes air exposure).\\

\begin{figure}
\includegraphics[width=12cm]{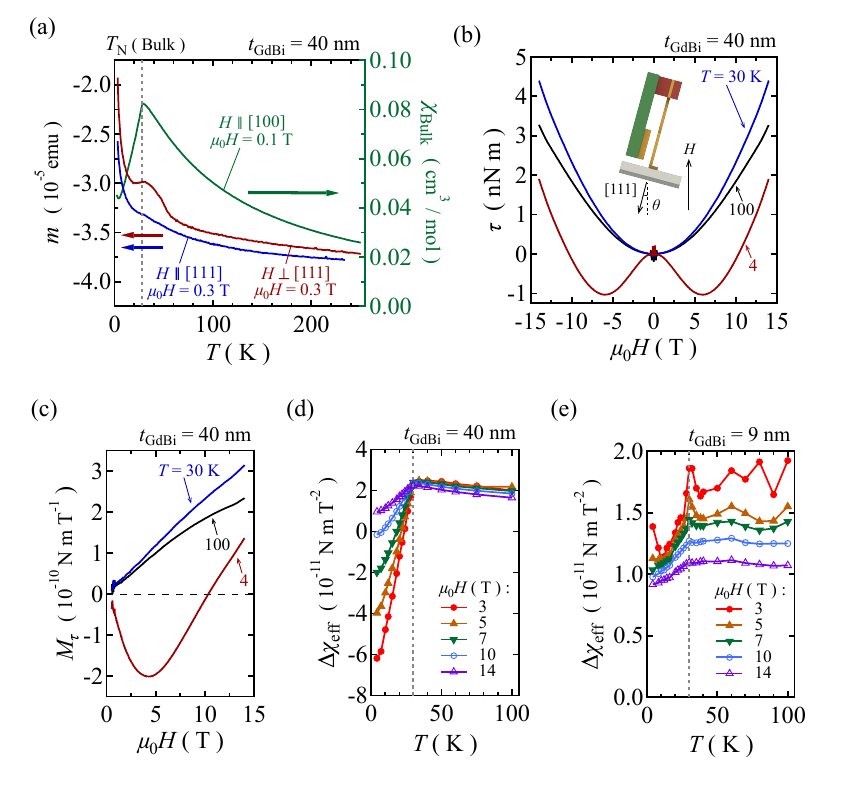}
\caption{\label{fig2}
AFM order in GdBi thin films. (a) Temperature dependence of the magnetic moment $m$ for $t_{\rm GdBi} = 40$ nm (red and blue curves, left axis) and magnetic susceptibility $\chi_{\rm Bulk}$ for a GdBi single crystal (green curve, right axis) measured by a SQUID magnetometer with field orientations as indicated. (b) Magnetic field dependence of magnetic torque $\tau$ of the GdBi thin film sample ($t_{\rm GdBi} = 40$ nm) grown on a 0.5-mm-thick BaF$_2$ substrate measured at different temperatures. The measurement geometry is shown in inset. (c) Magnetic field dependence of torque magnetization $M_{\tau}$ at different temperatures. (d)(e) Magnetic field dependence of $\Delta \chi_{\rm eff}$ calculated from $M_{\tau}$ for (d) $t_{\rm GdBi}$ = 40 nm and (e) $t_{\rm GdBi} = 9$ nm as a function of temperature (see text).
}
\end{figure}

\textbf{Magnetic characterization}

Figure \ref{fig2} (a) shows a comparison of the temperature $T$ dependence of magnetic susceptibility $\chi_{\rm Bulk}$ of a GdBi bulk single crystal and the magnetic moment $m$ of a 40 nm thick GdBi thin film measured by a commercial SQUID magnetometer.  As previously reported \cite{Duan2007,Ye2018}, the bulk crystal exhibits an AFM transition with Neel temperature $T_{N} = 28$ K as identified by the kink in $\chi_{\rm Bulk} (T)$ (measured here with the applied field $H$ parallel to [100]).  While for the thin film samples there is a relatively large background response arising from the substrate, buffer, and capping layers, there is a discernible peak for $T \sim T_{N}$ for $H$ perpendicular to [111].  For $H$ parallel to [111] this feature is largely suppressed, suggestive of the anisotropic magnetic susceptibility in the AFM phase.

In order to study the magnetic response of the films with higher resolution, we performed torque magnetometry experiments.  As shown in the inset of Fig$.$ \ref{fig2} (b), we mounted the sample directly to the end of a metal cantilever with a small angle $\theta \approx 15^{\circ}$ between $H$ and the sample normal.
The magnetic torque $\tau(H)$ is shown in Fig$.$ \ref{fig2} (b); at $T=100$ K we observe a quadratic response typical of a paramagnetic susceptibility.  This response is enhanced at $T=30$ K while a prominent dip at intermediate $H$ develops at the lowest $T = 4$ K. 
This is reminiscent of the W-shaped negative torque response originating from diamagnetism in superconducting states of high-$T_{\rm c}$ cuprates \cite{Wang2005}.

In Fig$.$ \ref{fig2} (c) we plot the corresponding torque magnetization $M_\tau \equiv \tau/\mu_0 H = M_{\rm plane} - M_{\rm norm}$, where $\mu_0$ is the vacuum permeability, $M_{\rm plane}$, and $M_{\rm norm}$ are the magnetization parallel and normal to the sample plane, respectively. Here the trend of an approximately linear susceptibility giving way to a strong non-linear response at low $T$ can also be observed.

The non-linear $M_{\tau}(H)$ at low $T$ can be naturally explained by the development of magnetic anisotropy upon entering the AFM phase as observed in the SQUID measurements. In the absence of magnetic field, the GdBi thin film forms antiferromagnetic domains, and the Gd spins point to symmetrically equivalent directions. Therefore the susceptibility is nearly isotropic for $H \sim 0$. However, application of magnetic field $H < 5$ T flops the spins due to anisotropic susceptibility of an antiferromagnet, and confine them within the sample plane. In this configuration, the spin susceptibility are anisotropic depending on magnetic field directions parallel or perpendicular to the sample plane. This generates the torque response as in Fig$.$ \ref{fig2} (b) acted by the tilted magnetic field. 

As the anisotropy of the magnetic susceptibility develops in the AFM state, it can be used to probe the Neel temperature of the films $T_{N}^{\textrm{film}}$.  We plot the observed anisotropy of effective magnetic susceptibility $\Delta \chi_{\rm eff} \equiv \chi_{\rm plane} - \chi_{\rm norm} = M_\tau/\mu_0 H$ in Fig$.$ \ref{fig2} (d) for different $H$ and identify $T_{N}^{\textrm{film}} = 30$ K.  The results for a thinner film with $t_{\textrm{GdBi}} = 9$ nm are shown in Fig$.$ \ref{fig2} (e) where a similar $T_{N}^{\textrm{film}}$ is observed.\\

\begin{figure}
\includegraphics[width=12cm]{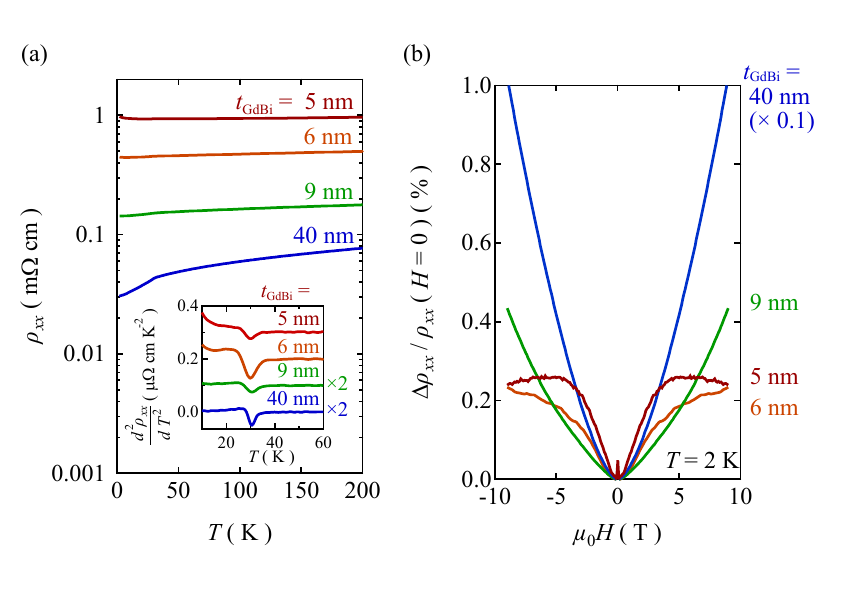}
\caption{\label{fig3}
Change of metallicity with thickness for GdBi thin films. (a) Temperature dependence of longitudinal resistivity $\rho_{xx} (T)$ for various $t_{\rm GdBi}$. Inset: Second derivative of $\rho_{xx} (T)$. (b) Magnetic field dependence of longitudinal resistivity $\rho_{xx} (H)$ measured at $T = 2$ K for various $t_{\rm GdBi}$. The data for $t_{\rm GdBi}=40$ nm are scaled by a factor of 0.1.
}
\end{figure}

\textbf{Electrical characterization}

The coupling between the conduction electron and localized magnetic moments of GdBi allows further characterization of the magnetic transition using electrical transport.  Figure \ref{fig3} (a) shows the $T$ dependence of longitudinal resistivity $\rho_{xx}(T)$ for different $t_{\rm GdBi}$.  For the thickest film with $t_{\rm GdBi} = 40$ nm we observe a metallic response for all $T$ with a kink in $\rho_{xx}(T)$ near $T=$ 30 K, the latter more clearly observed in the second derivative $d^2 \rho_{xx}/dT^2$ shown in the inset of Fig$.$ \ref{fig3} (a).  At all thicknesses down to $t_{\rm GdBi} = 5$ nm we observe this kink near $T=$ 30 K.  This approximately matches the observed $T_{N}^{\textrm{film}}$ shown in Figs. \ref{fig2} (d) and \ref{fig2} (e), suggesting that it is associated with the AFM transition.  Such a $\rho_{xx}(T)$ feature has been previously reported in GdBi bulk single crystals and ascribed to suppressed spin disorder scattering in the AFM phase \cite{Li1996}.  The observation of this feature in $d^2 \rho_{xx}/dT^2$ for all films down to $t_{\rm GdBi} = 5$ nm therefore suggests that $T_{\rm N}^{\textrm{film}}$ remains unchanged down to at least 5 crystallographic unit cells of GdBi.

As can be seen in Fig$.$ \ref{fig3} (a), the overall electrical response of the films changes from metallic to non-metallic with decreasing $t_{\rm GdBi}$; the low $T$ slope $d\rho_{xx}/dT$ changes from positive for $t_{\rm GdBi} = 40$ nm to negative for $t_{\rm GdBi} = 5$ nm.  This suggests a possible shift of the bulk band edges in the system with decreasing  $t_{\rm GdBi}$, which we address further below.

The magnetotransport response of this series of films at $T=2$ K is shown in Fig$.$ \ref{fig3} (b), where a non-saturating $(\rho_{xx}(H)-\rho_{xx}(H=0)) / \rho_{xx}(H=0)$ of approximately 10\% at $\mu_0 H=9$ T for the thickest films gives way to smaller, saturating behavior for thinner films.  Bulk single crystals of GdBi and other $RX$ systems have been reported to show extreme magnetoresistance (XMR) \cite{Ye2018,Kumar2016,FallahTafti2016,Tafti2016,Han2017} of similar form to that seen here for the thickest films, but with significantly larger amplitude in the bulk case.  As we discuss below, since XMR is influenced by band parameters including carrier density, mobility, and position of chemical potential \cite{Kumar2016}, the qualitative change of this response in the thin limit probes the evolution of these parameters as well as the band structure itself.

\begin{figure}
\includegraphics[width=16cm]{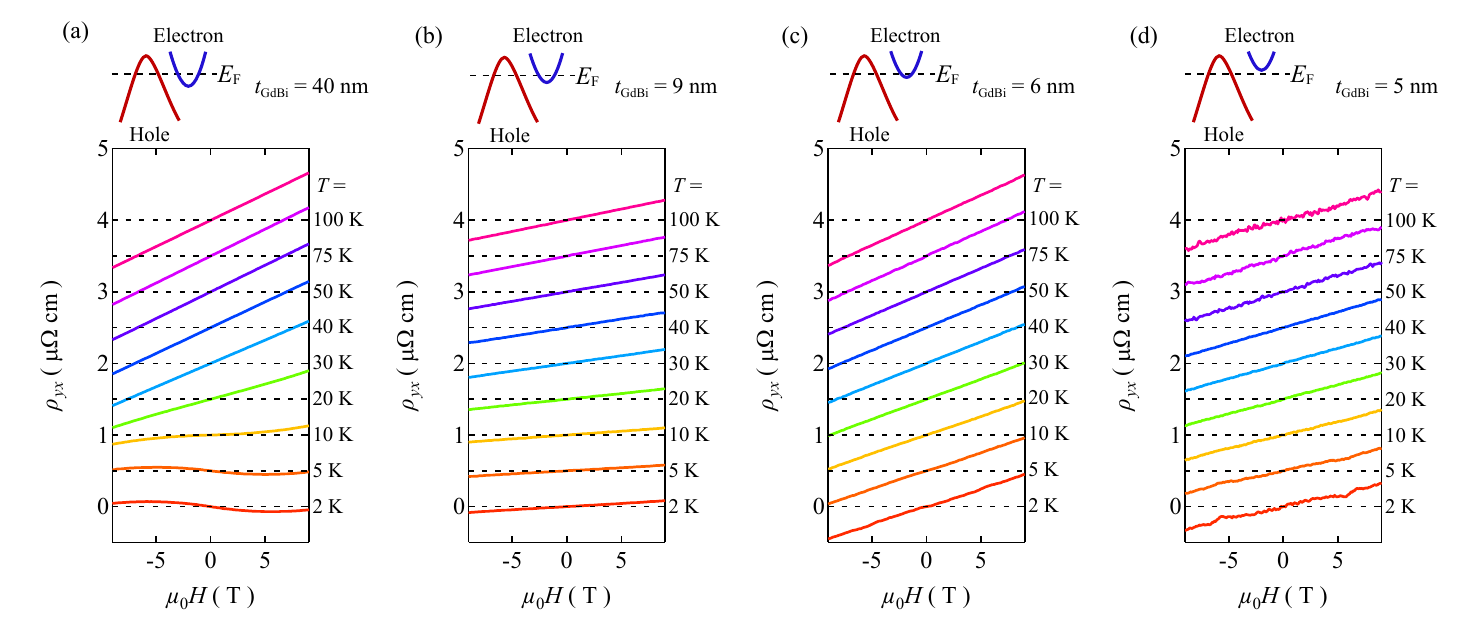}
\caption{\label{fig4}
Evolution of Hall effect as a function of GdBi film thickness. (a)(b)(c)(d) Transverse resistivity $\rho_{yx}$ as a function of magnetic field at various temperatures for (a) $t_{\rm GdBi} = 40$ nm, (b) $t_{\rm GdBi} = 9$ nm, (c) $t_{\rm GdBi} = 6$ nm, and (d) $t_{\rm GdBi} = 5$ nm.  The traces are offset vertically for clarity.  The upper inset in each panel is a schematic depiction of the overlapping bulk bands (see text).
}
\end{figure}

To further investigate the $t_{\textrm{GdBi}}$ dependence of transport, we measured the transverse (Hall) resistivity $\rho_{yx}(H)$ across a broad range of $T$.  As shown in Fig$.$ \ref{fig4} (a), for $t_{\textrm{GdBi}} = 40$ nm at $T=2$ K we observe a non-linear $\rho_{yx}(H)$ consistent with that expected from a semimetallic band structure.  With decreasing $t_{\textrm{GdBi}}$ (Figs. \ref{fig4} (b)-(d)), a linear response emerges suggestive of hole-like single-band transport.  At elevated $T$, $\rho_{yx}(H)$ for thicker films ($t_{\textrm{GdBi}} \geq 9$ nm) evolves towards a linear response whereas for thinner films ($t_{\textrm{GdBi}} \leq 6$ nm) it remains unchanged.\\

\textbf{First-principles Calculations}

The electrical transport and the torque data in GdBi films suggests suppressed metallicity in the thin limit, while long-range magnetic correlation is unaffected. If GdBi has a topologically non-trivial band structure, the consequence of broken time reversal symmetry may manifest itself as a topologically distinct phase such as a Chern insulating state characterized by its Chern number $C$ \cite{Haldane1988}, or antiferromagnetic topological insulator state with chiral edge modes on the step edges \cite{Mong2010}. Here we will show by theoretical calculations that GdBi has a topologically non-trivial band structure in both the bulk and thin film limit.

\begin{figure}
\includegraphics[width=15cm]{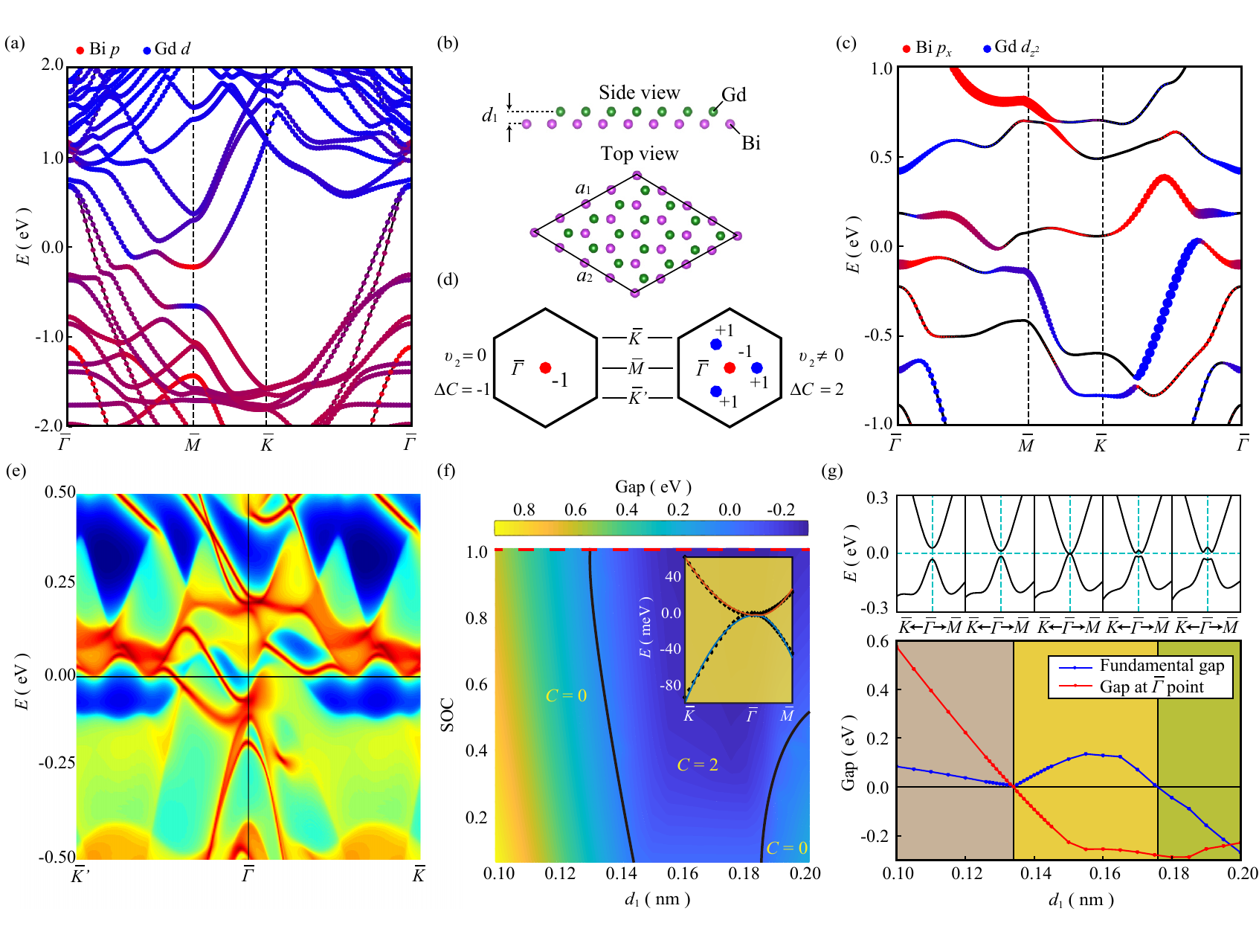}
\caption{\label{fig5}
First-principle calculations of GdBi band structures. (a) Band structure of bulk GdBi (111) with the spin-orbit coupling (SOC). (b) Illustrated crystal structure of the monolayer GdBi (111). $d_1$ is the distance between Bi and Gd layers and $a_1,a_2$ are the in-plane lattice constants. (c) Typical band structure of the monolayer GdBi (111). (d) Schematic Brillouin zone of the monolayer GdBi (111). Color of the circle denotes the change in the Chern number $\Delta C$ across the topological phase transition, contributed by each Dirac point: $-1$ for red and $+1$ for blue. (e) Edge states of monolayer GdBi (111) along Zigzag direction, exhibiting a pair of edge modes around the $\overline{\mathit{\Gamma}}$ point. (f) Topological phase diagram of monolayer GdBi (111) varies with inter-layer distance $d_1$ and SOC strength $\lambda$. The color represents the direct gap size and negative value indicates inverted band ordering. The inset shows the fitting results of $k\cdot p$ (solid line) to the first-principles calculations (empty circle) around the critical point $d_1=0.134$ nm. (g) Strain induced topological phase transition of the monolayer GdBi (111). The top and bottom panels respectively show the band evolution and energy gap as a function of $d_1$. In the bottom panel, the red line represents the direct gap at $\overline{\mathit{\Gamma}}$ point, and negative value means Chern insulator with $C=2$; the blue line represents the fundamental gap in the whole Brillouin Zone and the negative value denotes semi-metallic.
}
\end{figure}

Figure \ref{fig5} (a) shows the calculated band structure of bulk GdBi (111) with type-II AFM order in the hexagonal unit cell with spin-orbit coupling (SOC).
It exhibits an indirect negative band gap of about $-1$ eV between the Bi-derived valence band at $\mathit{\Gamma}$-point and the Gd-derived conduction band at $X$-points, forming a semimetal. From the systematic dependence of the calculated band structure as a function of the lattice constant, we confirmed that the bands are inverted at the $X$-points. We also calculated the surface states and find they connect the bulk valence bands and conduction bands. All calculations confirm that bulk GdBi has a non-trivial band topology corresponding to an AFM topological insulator gap degenerate with trivial bulk electronic states (see supplemental information).

In the extreme thin film limit, monolayer GdBi (111) consists of a pair of single atomic layers of Gd and Bi, separated by lattice spacing $d_1$ from each other (Fig$.$ \ref{fig5} (b)). To determine the topology of GdBi (111) in the 2D limit, the band structure of monolayer GdBi (111) slab was calculated using $d_1 = 0.182$ nm (the bulk value) as shown in Fig$.$ \ref{fig5} (c). We observe that the size of the negative indirect band gap is nearly lifted owing to the quantum confinement. The orbital-projection analysis reveals band inversion between the Gd $d_{z^2}$ orbital and Bi $p_x$ orbital at the $\overline{\mathit{\Gamma}}$ point, implying a nontrivial band topology. \\

\textbf{Symmetry Analysis and Edge Modes}

In order to determine the topological properties relevant to monolayer GdBi (111), we performed symmetry analysis of the low energy band structure. Monolayer GdBi (111) has 3-fold rotation symmetry $C_3$. Due to the magnetic order, both time-reversal symmetry $T$ and mirror symmetry with the mirror plane perpendicular to $y$-direction $M_y$ are broken, but the combined $TM_y$ is preserved. To characterize the low-energy properties near $\overline{\mathit{\Gamma}}$, we use the little group containing $C_3$ and $TM_y$ symmetries to build a $k\cdot p$ Hamiltonian 
\begin{eqnarray}
H \left( k_x,k_y \right) = m_1\sigma_z + m_2 (k_x^2+k_y^2)\sigma_z  + v_1 \left( k_x \sigma_y - k_y \sigma_x \right) \nonumber \\
+ v_2 \left(2k_x k_y \sigma_y -\left( k_x^2-k_y^2 \right) \sigma_x \right) + v_3 \left( k_x^2+k_y^2 \right)I,
\end{eqnarray}
where $\sigma$ is the pseudo-spin representing the conduction and valence bands, $m_1$ and $m_2$ are mass parameters, $k_x$ and $k_y$ are the crystal momenta, and $v_1, v_2$, and $v_3$ are the velocity parameters. The $v_3$ term gives the same energy shift for both conduction and valence bandsand thus will not affect the topological properties. We therefore set $v_3=0$ in the following.

For the simplest case with $m_2=v_2=0$, the $k\cdot p$ Hamiltonian reduces to the typical massive Dirac Hamiltonian. For $m_1 \neq 0$, a gap will open at  $\overline{\mathit{\Gamma}}$ (red circles in Fig$.$ \ref{fig5} (d)). Across this band inversion, a topological phase transition occurs with Chern number changed by $\Delta C = -1$.  In the case of small $v_2 \ne 0$, Dirac cones appear at four different points: one at $\overline{\mathit{\Gamma}}$ and the other three at equivalent points along the $\overline{\mathit{\Gamma}}-\overline{M}$ lines (blue circles in Fig$.$ \ref{fig5} (d)). As $m_1$ changes from negative to positive, the total Chern number now changes by $\Delta C=2$.  For $v_2>>v_1$, as $v_2$ increases the three Dirac points along the $\overline{\mathit{\Gamma}}-\overline{M}$ line converge at $\overline{\mathit{\Gamma}}$ and transform into a quadratic band touching. As in the case of small $v_2$, $\Delta C=2$ when $m_{1}$ changes sign.  For $m_2\ne0$, $m_2 (k_x^2+k_y^2)=0$ at $\overline{\mathit{\Gamma}}$ and $m_2$ will not affect the gap closing or reopening there.  However, $m_{2}$ does affect the gap along the $\overline{\mathit{\Gamma}}-\overline{M}$ lines. By fine-tuning $m_2$, we can also realize an intermediate phase with $C=-1$ or $3$ between $C=0$ and $C=2$ phases; the parameter regime of the intermediate phase is $\delta=\left|m_2\right|(\frac{v_1}{v_2})^2$. 

To confirm the Chern insulating state with $C=2$, we show in Fig$.$ \ref{fig5} (e) the edge states along Zigzag direction, which are calculated in \textit{ab-initio} tight-binding models with all the parameters fitted from the first-principles calculations through Wannier90 \cite{Wannier90}. It shows two chiral edge modes connecting the valence and the conduction bands around the $\overline{\mathit{\Gamma}}$ point. These modes always appear as a pair, reflecting the topological band character $C = 2$. \\

\textbf{Calculations of Topological Phase Transition}

We performed systematic calculations for the topological phase diagram for monolayer GdBi by varying the interlayer distance $d_1$ and the SOC strength $\lambda$. As shown in the Fig. \ref{fig5} (f), the phase diagram contains only phases with $C=0$ and $C=2$. As shown in the top panel of Fig. \ref{fig5} (g), the conduction band and valence band have quadratic touching at the band inversion (similar behavior happens when we vary SOC strength $\lambda$ as shown in Supplementary Information), which indicates a high-order topological phase transition.  Fitting all the parameters in the $k\cdot p$ model to the band structures around the critical distance $d=1.34$  (inset of Fig. \ref{fig5} (g)), we find $v_1=-0.3$, $v_2=37.2$, $v_3=-7.1$, and  $|m_2| < 0.1$.  This is consistent with the phase diagram obtained by first-principles calculations (the intermediate phase $\delta=|m_2|(\frac{v_1}{v_2})^2\sim 10^{-5}$ is extremely small). In Fig$.$ \ref{fig5} (g), we show the evolution of the energy gap $\Delta$ at the $\overline{\mathit{\Gamma}}$ point as a function of $d_1$. $\Delta$ monotonically decreases when $d_1$ increases for $d_1 < 0.134$ nm. At $d_1=0.134$, $\Delta=0$ and the gap reopens for $d_1 > 0.134$ nm indicating a topological phase transition. \\

\textbf{Discussion}

The GdBi thin films studied here appear to retain the magnetic properties of bulk crystals down to $t_{\textrm{GdBi}} = 5$ nm, suggesting that the symmetry breaking AFM order of these materials is not significantly altered.  The transport response, on the other hand, does evolve significantly on decreasing thickness particularly across $t_{\textrm{GdBi}} = 9$ nm.  By quantitatively analyzing the magnetotransport results, we can connect this behavior with that reported for bulk single crystals as well as that predicted by our theoretical calculations in the thin limit.

Starting with the thickest film with $t_{\textrm{GdBi}} = 40$ nm, we expect a minimal role of quantum confinement and therefore transport behavior similar to bulk single crystal materials.  While we do observe a non-saturating magnetoresistivity (Fig$.$ \ref{fig3} (b)), the overall magnitude of this response is significantly smaller than the XMR behavior reported in bulk single crystals \cite{Tafti2016}.  This can be understood in terms of the compensation model for XMR which requires a balance of density and mobility of the conduction and valence bands \cite{Ali2014,Shekhar2015}.  While compensation has been observed in bulk $RX$ systems \cite{Ye2018,Han2017,Zeng2016a,FallahTafti2016,Guo2016,He2016,Wakeham2016}, two band analysis (see supplementary materials) of the magnetotransport results for our $t_{\textrm{GdBi}} = 40$ nm film (Fig$.$ \ref{fig4} (a)) yields $n_{\rm e} = 2.5 \times 10^{20}$ cm$^{-2}$, $n_{\rm h} = 3.0 \times 10^{20}$ cm$^{-2}$, $\mu_{\rm e} = 389$ cm$^2$/Vs, $\mu_{\rm h} = 349$ cm$^2$/Vs for $t_{\rm GdBi} = 40$ nm at $T = 2$ K, where $n_{\rm e}$, $n_{\rm h}$, $\mu_{\rm e}$, and $\mu_{\rm h}$ are the electron density, the total hole density, the electron mobility and the hole mobility, respectively.  While the carrier densities are similar to those observed in bulk GdBi, along with reduced overall mobility they are detuned enough from perfect compensation to explain the significantly reduced XMR response \cite{Wang2015,Tafti2017,Sun2016}. The origin of this difference of parameters could arise from a number of sources including charge transfer from the substrate, defect chemistry differences in bulk and thin film synthesis, or epitaxial strain.

Upon decreasing $t_{\textrm{GdBi}}$, the compensation is further removed such that by $t_{\textrm{GdBi}} = 6$ nm the $T = 2$ K Hall effect is captured by a purely hole-like $\rho_{yx}(H)$ (see Fig$.$ \ref{fig4} (c)). A plausible explanation for this is a confinement induced \cite{Li2015,Chatterjee2019} upward shifting of the electron band.  This scenario is depicted schematically in the upper insets of Fig$.$ \ref{fig4} (a)-(d) and corresponds to a decrease of the semimetallic band overlap at $E_{\rm F}$.  This is further consistent with the change from metallic to mildly insulating behavior in $\rho_{xx}(T)$ shown in Fig$.$ \ref{fig3} (a): here the hole-like bands remain metallic but the upwards shifted electron band has a parallel thermally activated conductivity with a relatively minor contribution to the off-diagonal response.  Stabilizing films with further reduced $t_{\textrm{GdBi}}$ would then be expected to further enhance quantum confinement toward the realization of $C=2$ Chern insulator state as predicted in our \textit{ab-initio} calculations.  We note that in the ultrathin limit Anderson localization is also relevant \cite{Anderson1958}; this will act to eventually gap out the non-trivial electronic states and, in the case of a time-reversal symmetry breaking by canting of the AFM order, potentially isolate chiral modes across a broad energy range \cite{Nomura2011}. 

Analyzing our theoretical calculations offers insight in to the underlying mechanism by which monolayer GdBi may realize a $C=2$ phase.  This appears to be a result of the collapse of the intermediate phase corresponding to more conventional $\Delta C=1$ transitions found when $\left|m_2\right|$ is small and $v_2$ is very large (and further that this is the physically relevant regime for monolayer GdBi).  This approach to realizing $C=2$ Chern insulating state is general and may be relevant to the recently proposed $C=2$ state in the Dice lattice with $C_{3v}$ symmetry \cite{Wang2011} and also to the topological crystalline insulator SnTe with the mirror Chern number $2$ \cite{Hsieh2012}.  Our further demonstration that this phase is sensitive to the layer spacing parameter $d_{1}$ suggest the opportunity to engineer band topology using lattice strain, which can be controlled either by external pressure or by epitaxial strain \cite{liu2014manipulating}. Therefore monolayer GdBi (111) is an attractive system to realize a tunable $C=2$ Chern insulator state. \\

\textbf{Conclusion}

We have reported the structural, magnetic, and transport properties of the correlated topological insulator candidate GdBi thin films.  We find bringing materials to the thin film limit preserves the antiferromagentic properties of bulk single crystals but with modified electrical properties consistent with a lifting of the semimetallic band overlap.  Together with our first principles calculations that show the band topology is preserved in the monolayer limit, this demonstrates that GdBi and other magnetic $RX$ systems in thin film form are candidates for hosting intrinsic correlated topological phases, including antiferromagnetic TIs \cite{Mong2010}, and $C=2$ Chern insulating phase tunable by strain in the monolayer limit \cite{Yu2010,Haldane1988,Chang2013,Checkelsky2014}.  We also note that the films here have a $(111)$ orientation that has thus far not been stabilized for spectroscopic studies in bulk single crystals \cite{Zeng2016a,Nayak2017,Kuroda2018,Lou2017,Feng2018,Wu2017,Alidoust2016}.  Theoretical calculations for $R$Bi predict that three surface Dirac cones are each separately projected on to distinct points in the Brillouin zones on the $(111)$ surface, but that two are degenerate and obscured by bulk bands for the (currently available) $(001)$ surface \cite{Zeng2015}.  Therefore, the high quality $(111)$-oriented films reported here even in the thick limit may serve an important role for improved spectroscopic characterization.  Finally, following the methodology presented herein, we expect that other members of the $RX$ family with more complex magnetic phase diagrams including CeBi \cite{Rossat-Mignod1983} and HoBi \cite{Hulliger1984,Fente2013} should also be readily synthesized, allowing for exploration of a wide variety of novel symmetry broken phases in this new class of topological electronic materials.  \\

\appendix

\textbf{Methods}

\textbf{Sample fabrication}

Thin film samples were deposited on BaF$_2$ $(111)$ substrates using an MBE system with base pressure $4 \times 10^{-8}$ Pa. Prior to the BaF$_2$ buffer layer deposition, the BaF$_2$ substrate was annealed at $T = 450$ $^\circ$C for 90 minutes. The BaF$_2$ buffer was then grown at $T = 765$ $^\circ$C for 150 minutes at a growth rate of 1.33 nm/min. Next, the growth stage was brought to $T=400$ $^\circ$C and GdBi was grown at a rate of 0.22 nm/min. The ratio between the Gd and Bi flux was $1:4.34$. Finally an epitaxial BaF$_2$ cap layer was grown at $T = 765$ $^\circ$C for 60 minutes at a growth rate of 1.33 nm/min. Samples were transferred into a glovebox directly after the growth without air exposure to prevent sample degradation before subsequent processing. \\

\textbf{Magnetometry using superconducting quantum interference device}

We characterized the magnetization of GdBi thin films using a commercial magnetometer with a superconducting quantum interference device (SQUID). To prevent degradation of the films, samples were \textit{ex-situ} coated with an additional 30 nm of AlN layer at 120$^\circ$C by atomic layer deposition at a rate of 0.10 nm per cycle. Samples were immersed in acetone prior to deposition in order to prevent degradation. We confirmed that the magnetic properties of the samples stayed unchanged during the measurements.\\

\textbf{Torque magnetometry} 

After deposition of the GdBi thin films, samples were transferred into an inert environment from the MBE chamber and then attached to a 10 {\textmu}m-thick Au torque cantilever. A cryogenic insert fitted with a small vacuum chamber was loaded into the inert environment, and the magnetometer was placed in the vacuum chamber. Before extracting the cryogenic insert from the inert environment and loading it into a cryostat, the vacuum chamber was evacuated using an external pump. The vacuum chamber was kept under vacuum during the measurements to prevent degradation of the film.  The deflection of the cantilever in magnetic field was measured via capacitance between the flexible cantilever and a fixed electrode using a capacitance bridge.\\
\\

\textbf{Electrical transport measurement}

Electrical transport properties of the GdBi thin films were characterized using the Van der Pauw geometry in a commercial cryostat. We applied a non-aqueous liquid electrolyte ($N$,$N$-Diethyl-$N$-methyl-$N$-(2-methoxyethyl)ammonium bis(trifluoromethanesulfonyl)imide) and made electrical contacts to the thin films in an inert environment to prevent degradation of the thin films after direct transfer from the MBE chamber. There was no measurable changes in the electrical properties after deposition of the liquid. \\ 

\textbf{First-principles calculations}

We performed first-principles calculations of the electronic structures of bulk and ultrathin GdBi (111) in the type-II AFM phase. The calculations are performed in the framework of density functional theory (DFT) as implemented in the Vienna \emph{ab initio} simulation package (VASP) \cite{VASP} by the Perdew-Burke-Ernzerhof type of generalized gradient approximation (GGA) \cite{PBE} and the projector augmented wave (PAW) method \cite{PAW} with cutoff energy as 400 eV. The band structures were calculated using the Monkhorst-pack ($6 \times 6 \times 6$) and ($6 \times 6 \times 1$) $k$-mesh to sample the Brillouin zones of the bulk and thin films. The Hubbard correlation term $U$ was introduced into the density functional theory framework (DFT+U) in order to account for the electron-electron correlation. The experimental lattice constant of bulk GdBi ($a = 0.632$ nm) was used for bulk calculations. For the value of $U$, we found that the calculated lattice constant reproduces the experimental value when $U$ around 12 eV. Therefore we set $U=12$ eV for the calculations.  The spin orientation is set parallel to [111] in a type-II AFM order. All the parameters are tested to confirm the calculations converged, and the calculated bulk band structure is consistent with previous calculations \cite{Larson2006,Masrour2014}.

\bibliography{GdBiThinFilm_v04}

\vspace{10 mm}

\textbf{Acknowledgments}  

We are grateful to L. Fu and C. Fang for fruitful discussions. This research was funded, in part, by the Gordon and Betty Moore Foundation EPiQS Initiative, Grant No. GBMF3848 to J.G.C. and  ARO Grant No. W911NF-16-1-0034. J.G.C. acknowledges support from the Bose Fellows Program at MIT.  M. Hu and J. Liu acknowledge financial support from the Hong Kong Research Grants Council (Project No. ECS26302118). \\

\end{document}